\def\tr{{\text{tr}}\,}

\def\Im{{\text{Im}}\,}
\def\Re{{\text{Re}}\,}
\def\kF{k_{\text{F}}}
\def\vF{v_{\text{F}}}
\def\NF{N_{\text{F}}}
\def\me{m_{\text{e}}}
\def\ne{n_{\text{e}}}
\def\epsilonF{\epsilon_{\text F}}
\def\TF{T_{\text{F}}}
\def\Gammat{\Gamma_{\text{t}}}

\def\sgn{{\text{sgn\,}}}
\def\be{\begin{equation}}
\def\ee{\end{equation}}
\def\bea{\begin{eqnarray}}
\def\eea{\end{eqnarray}}
\def\bse{\begin{subequations}}
\def\ese{\end{subequations}}

\documentclass[prb,twocolumn,showpacs,amsmath,amssymb,eqsecnum,preprintnumbers]{revtex4}
\usepackage{graphicx}
\usepackage{dcolumn}
\usepackage{bm}
\begin{document}
\title{Electronic relaxation rates in metallic ferromagnets}
\author{S. Bharadwaj$^1$, D. Belitz$^{1,2}$, and T.R. Kirkpatrick$^3$}
\affiliation{$^{1}$ Department of Physics and Institute of Theoretical Science,
                    University of Oregon, Eugene, OR 97403, USA\\
                   $^{2}$ Materials Science Institute, University of Oregon, Eugene,
                    OR 97403, USA\\
                  $^{3}$ Institute for Physical Science and Technology,
                    and Department of Physics, University of Maryland, College Park,
                    MD 20742, USA
            }
\date{\today}

\begin{abstract}
We show that the magnon-exchange contribution to the single-particle and transport relaxation rates in ferromagnetic 
metals, which determine the thermal and electrical conductivity, respectively, at asymptotically low temperature does 
not obey a power law as previously thought, but rather shows an exponential temperature dependence. The reason 
is the splitting of the conduction band that inevitably results from a nonzero magnetization. At higher temperatures 
there is a sizable temperature window where the transport rate shows a $T^2$ temperature dependence, in accord 
with prior results. This window is separated from the asymptotic regime by a temperature scale that is estimated to 
range from tens of mK to tens of K for typical ferromagnets. We motivate and derive a very general effective theory 
for metallic magnets that we then use to derive these results. Comparisons with existing experiments are discussed, 
and predictions for future experiments at low temperatures are made.
\end{abstract}

\pacs{72.10.Di; 72.15.Lh; 75.30.Ds}

\maketitle

\section{Introduction}
\label{sec:I}

Electronic relaxation rates contain important information about the excitations in a metallic system.
The single-particle relaxation rate, $1/\tau$, determines the lifetime of quasi-particles as well as the
thermal conductivity $\kappa = \vF^2\,c_V\tau/3$; the transport relaxation rate, $1/\tau_{\text{tr}}$, 
the electrical conductivity via the Drude formula $\sigma = \ne e^2\tau_{\text{tr}}/\me$. Here $\me$ 
and $\ne$ are the conduction electron effective mass and number density, respectively, $\vF$ is the 
Fermi velocity, and $c_V$ is the specific heat. There are various contributions to these relaxation 
rates, including those from the scattering of electrons by propagating, or particle-like, excitations. 
For instance, the coupling of longitudinal phonons to conduction electrons leads to the well-known 
Bloch $T^5$-behavior of the electrical resistivity; the corresponding effect in the single-particle
relaxation rate is a $T^3$-law.\cite{Ziman_1960} In magnetically ordered phases, the coupling of 
the conduction electrons to any magnetic Goldstone modes contributes to the relaxation rates. In 
isotropic Heisenberg ferromagnets, the Goldstone modes are the ferromagnons with a 
frequency-momentum relation $\omega \sim k^2$. They have been found to contribute
a $T^2$ term to the transport relaxation rate.\cite{Moriya_1985, T_squared_history_footnote}
In helimagnets,\cite{Dzyaloshinsky_1958, Moriya_1960} which have a helically modulated magnetic 
ground state, the corresponding Goldstone mode (the helimagnon) has been shown to lead to 
a term in the electrical resistivity that is proportional to $T^{5/2}$ in the low-temperature 
limit.\cite{Belitz_Kirkpatrick_Rosch_2006b, Kirkpatrick_Belitz_Saha_2008a, Kirkpatrick_Belitz_Saha_2008b}
In antiferromagnets, the corresponding contribution is known to be proportional to $T^3$.\cite{Ueda_1977}
These results all hold for three-dimensional systems, which is the only physical dimension in which
long-range magnetic order exists. For later reference we note, however, that the various power
laws quoted above are dimensionality dependent. For instance, in a generic dimension $d>2$
the contribution from ferromagnons to the resistivity is proportional to $T^{(d+1)/2}$.

In addition to the scattering by propagating excitations, there are contributions to the transport
coefficients due to excitations with a continuous spectrum. The best known example is the one
due to the Coulomb interaction between the electrons. In simple metals it leads to a $T^2$
contribution to both the single-particle rate and the transport rate, i.e., a lower power
than the phonon contribution. However, since the relevant energy scale is the Fermi energy 
$\epsilonF$ or Fermi temperature $\TF$ (we use units such that $\hbar = k_{\text{B}} = 1$), which 
is much larger than the Debye temperature, this dominates the phonon contribution only at very low 
temperatures.\cite{Landau_Lifshitz_X_1981} In metals that display ferromagnetism the latter 
statement is not necessarily true, due to Fermi surfaces that consist of multiple sheets, and the
issue of both the temperature dependence and the prefactor of the Coulomb contribution to the 
electrical resistivity is complicated. These are old
questions\cite{Peierls_1930, Wilson_1954, Ziman_1960} that recently have been revisited in the context
of quantum criticality and exotic metals.\cite{Caprara_et_al_2007, Pal_Yudson_Maslov_2012} 
Another example is the scattering of electrons in ferromagnets by both longitudinal magnetization fluctuations
and the so-called Stoner excitations in the transverse channel.\cite{Moriya_1985} The latter are dissipative, non-hydrodynamic
transverse excitations in addition to the propagating spin waves. In a random-phase approximation, the contribution to the
resistivity from these dissipative excitations, both longitudinal and transverse, was shown in 
Ref.\ \onlinecite{Ueda_Moriya_1975} to result in a $T^2$ behavior with a prefactor that is inversely
proportional to the magnetization. This is qualitatively the same behavior these authors found for
the scattering by magnons, and it agrees roughly with the trend observed in Fe, Co, and Ni.\cite{Campbell_Fert_1982} 
As we will see, this conclusion, as far as the magnons is concerned, is
true only in a temperature window, but not at asymptotically low temperature. It should be stressed, however,
that this similarity is somewhat accidental and approximation dependent even in the regime where it holds. 
For instance, the power law of the magnon contribution is
dimensionality dependent, as mentioned above, while the contribution from the dissipative excitations
is not. Also, the prefactor of the former is essentially determined by the dispersion relation of the magnons,
which is governed by very general principles, whereas the latter is dependent on many non-universal
details. Nevertheless, the fact that various contributions of very different nature to the relaxation
rates show a $T^2$ temperature dependence makes the interpretation of the experimentally observed
$T^2$ behavior of the electrical resistivity in many ferromagnetic materials difficult.\cite{Campbell_Fert_1982}
At the same time, the electrical resistivity is a basic physical property that is very useful,
for instance, for tracking and identifying magnetic phase transitions,\cite{Taufour_et_al_2010,
Steppke_et_al_2013, Kotegawa_et_al_2011, Huang_et_al_2013, Aoki_et_al_2011b} and establishing its
behavior in the ferromagnetic phase as a benchmark is important.

In this paper we focus on the magnon contribution to the relaxation rates in ferromagnets and show that
for this process the established result is qualitatively incorrect at asymptotically low temperatures; instead 
of a $T^2$ temperature dependence, the magnon contributions to both the electrical resistivity and the 
thermal resistivity display an exponential behavior. A problem with the established ferromagnetic result was first noted in
Ref.\ \onlinecite{Ho_et_al_2010}, which showed that the results for the helimagnetic and ferromagnetic cases
are not mutually consistent: If one considers the ferromagnetic limit of the helimagnetic
ground state, by letting the wavelength of the helix go to infinity, one finds that the leading
contribution to the relaxation rate, which would yield a power law, vanishes. What is left
behind is an exponential behavior of the form\cite{T_squared_footnote}
\be
1/\tau_{\text{tr}} \propto (T^2/\lambda)\,\exp(-T_0/T)\ , 
\label{eq:1.1}
\ee
where the temperature scale $T_0$ depends on the conduction band splitting or ``Stoner gap'' $\lambda$ or, 
equivalently, the magnetization,\cite{Stoner_gap_footnote} and on the Fermi energy $\epsilonF$. 
This result is surprising, given that the relaxation rates due to magnetic Goldstone modes in both
helimagnets and antiferromagnets show a power-law behavior. 
The purpose of this paper is to discuss this problem, and to elaborate on the brief remarks 
that were given in Appendix D of Ref.\ \onlinecite{Ho_et_al_2010}. We will show that the asymptotic 
low-temperature behavior of both the transport relaxation rates due to magnons is indeed exponential
of the form shown in Eq.\ (\ref{eq:1.1}), with $T_0 \approx D\kF^2(\lambda/\epsilonF)^2$ with $D$ the
spin-wave stiffness, which itself depends on $\lambda$, and $\kF$ and $\epsilonF$ the Fermi wave
number and Fermi energy, respectively. This result holds
in an asymptotic regime defined by $T \ll T_0$. However, in a sizable
pre-asymptotic temperature window given by $T_0 \ll T \ll D\kF^2$ one recovers
the $T^2$ behavior found previously. The reason for the exponential asymptotic result
is the fact that, in a ferromagnet, the Goldstone modes are purely transverse, and therefore couple
only quasi-particles in different Stoner bands. The effective electron-electron-interaction due to
ferromagnon exchange therefore describes purely inter-Stoner-band scattering, which leads to
an activated process. In contrast, in helimagnets and antiferromagnets there is an intra-Stoner-band
coupling which leads to a power law. This vanishes as the characteristic wave number of the
magnetic order goes to zero in the ferromagnetic limit.

These results are valid for all metallic ferromagnets, whether or not the magnetism is caused
by the conduction electrons themselves or by localized electrons in a different band. We will 
refer to such systems as ``itinerant ferromagnets'' and ``localized-moment ferromagnetes'',
respectively. In the main body of the paper we will consider a very general model that 
does not depend on which of these two cases is realized, and that uses only very general 
properties of ferromagnets that follow from symmetry arguments. A more specific Stoner-type
model for the case of itinerant ferromagnets is considered in an appendix.

This paper is organized as follows. In Sec. \ref{sec:II} we derive an effective action that
describes an effective electron-electron interaction due to the exchange of ferromagnons.
The effective action is valid for calculating relaxation rates to first order in the magnon
propagator, and it holds for both itinerant and localized-moment ferromagnets. In Sec.\ \ref{sec:III} we use
this model to calculate the single-particle relaxation time, and in Sec.\ \ref{sec:IV} we calculate 
the transport relaxation time, and hence the electrical conductivity, by evaluating the pertinent Kubo 
formula in an approximation that is equivalent to the Boltzmann equation. In Sec.\ \ref{sec:V} 
we discuss our results. In Appendix A we recall the Stoner-Moriya mean-field treatment of itinerant ferromagnets. 
In Appendix B we recall the cases of electron-electron and electron-phonon scattering in non-magnetic metals,
and cast them in a language that illustrates why our general method works even in the case of itinerant ferromagnets.

\section{Effective action}
\label{sec:II}

In this section we derive and motivate an effective action that is suitable for calculating the effects of
long-range ferromagnetic order, and the associated Goldstone modes, on the electronic
relaxation rates in a metallic ferromagnet. 

\subsection{Coupling of magnetic fluctuations to conduction electrons}
\label{subsec:II.A}

Let $S_0[{\bar\psi},\psi]$ be an action for
conduction electrons in terms of  fermionic spinor fields ${\bar\psi} = ({\bar\psi}_{\uparrow},{\bar\psi}_{\downarrow})$
and $\psi = (\psi_{\uparrow},\psi_{\downarrow})$ that depend on a spin projection index
$\sigma = (\uparrow,\downarrow) \equiv (+,-)$. The electronic spin density is given
by
\be
{\bm n}_{\text{s}}(x) = \sum_{\sigma,\sigma'} {\bar\psi}_{\sigma}(x)\,{\bm \sigma}_{\sigma\sigma'}\,\psi_{\sigma'}(x)\ .
\label{eq:2.1}
\ee
Here ${\bm\sigma} = (\sigma^1,\sigma^2,\sigma^3)$ denotes the Pauli matrices, and 
$x = ({\bm x},\tau)$ comprises the real-space position ${\bm x}$ and the imaginary-time
variable $\tau$. Now assume that the conduction electrons are subject to a 
magnetization ${\bm M}(x)$ of unspecified origin. The magnetization will act as an effective magnetic field
that couples to the conduction electrons via a Zeeman term. The action then reads
\be
S[{\bar\psi},\psi] = S_0[{\bar\psi},\psi] + \Gammat \int dx\ {\bm M}(x)\cdot{\bm n}_{\text{s}}(x)\ ,
\label{eq:2.1'}
\ee
with $\Gammat$ a coupling constant that dimensionally is an energy times a volume, or an inverse
density of states. In a ferromagnetic state, the magnetization has a nonzero average value that we
assume to be in the 3-direction, $\langle M_i(x)\rangle = \delta_{i3}\,m$. In a mean-field approximation
that replaces ${\bm M}$ by its average value the action then takes the form
\bse
\label{eqs:2.2}
\be
S_{\lambda}[{\bar\psi},\psi] = S_0[{\bar\psi},\psi] + \lambda \int dx\ n_{\text{s},3}(x)\ ,
\label{eq:2.2a}
\ee
where $\lambda = \Gammat m$ is directly proportional to the average magnetization.
Here we have chosen the sign of the action such that the partition function is given by
\be
Z_{\lambda} = \int D[{\bar\psi},\psi]\ e^{S_{\lambda}[{\bar\psi},\psi]}\ .
\label{eq:2.2b}
\ee
\ese
$\lambda$ splits the conduction band into two sub-bands, one for each spin projection.
We will refer to $\lambda$ as the Stoner gap,\cite{Stoner_gap_footnote} but we emphasize that the physical
situation we are considering is much more general than the one considered in the
Stoner model.\cite{Stoner_1938}  In particular, we do not necessarily assume that
the conduction electrons themselves are the source of the magnetization.

Now consider fluctuations $\delta{\bm M}$ of the magnetization. The action, Eq.\ (\ref{eq:2.1'}), then reads
\bse
\label{eqs:2.4}
\be
S[{\bar\psi},\psi] = S_{\lambda}[{\bar\psi},\psi] + \Gamma_{\text{t}} \int dx\ \delta{\bm M}(x)\cdot{\bm n}_{\text{s}}(x)\ .
\label{eq:2.4a}
\ee
In addition we need an action that governs $\delta{\bm M}$. If the latter is to describe the fluctuations of the physical
magnetization, then this must be 
\be
S_{\text{fluct}}[\delta {\bm M}]  = \frac{-1}{2} \int dx\,dy\ \delta M_{\text{s},i}(x)\,\chi^{-1}_{ij}(x,y)\,\delta M_{\text{s},j}(y)\ ,
\label{eq:2.4b}
\ee
\ese
where $\chi_{ij}(x,y)$ is the physical magnetic susceptibility. In a ferromagnetic phase, the
transverse ($i,j=1,2$ with our choice for the magnetization direction) components of $\chi_{ij}$
contain the ferromagnons, which are the Goldstone modes associated with the ferromagnetic
order. The transverse part of $\chi_{ij}$ is thus singular in the limit of small frequencies and
wave numbers. Adding Eqs.\ (\ref{eq:2.4a}) and (\ref{eq:2.4b}), and integrating out $\delta{\bm M}$, 
we obtain a purely electronic effective action
\bse
\label{eqs:2.5}
\be
S_{\text{eff}}[{\bar\psi},\psi] = S_{\lambda}[{\bar\psi},\psi] + S_{\text{ex}}[{\bar\psi},\psi]\ ,
\label{eq:2.5a}
\ee
with
\be
S_{\text{ex}}[{\bar\psi},\psi] = \frac{\Gamma_{\text{t}}^2}{2}\int dx dy\, \delta n_{\text{s},i}(x)\,\chi^{}_{ij}(x,y)\,\delta n_{\text{s},j}(y) \ .
\label{eq:2.5b}
\ee
\ese
If we use only the singular,
transverse, part of $\chi_{ij}$, then $S_{\text{ex}}$ describes an effective electron-electron
interaction mediated by an exchange of magnons.\cite{chi_L_footnote}

\subsection{Effective action}
\label{subsec:II.B}

In order to make the effective action given by Eqs.\ (\ref{eqs:2.5}) suitable for explicit calculations, we
now specify $S_0$ and $\chi_{ij}$. The former in principle describes interacting electrons in a conduction
band. However, the electron-electron interaction is not of any qualitative importance for our purposes,
and we therefore take $S_0$ to describe noninteracting electrons with an energy-momentum relation
$\epsilon_{\bm k}$. We denote the chemical potential by $\mu$, and define $\xi_{\bm k} = \epsilon_{\bm k} - \mu$.
$S_{\lambda}$ then reads
\bse
\label{eqs:2.6}
\be
S_{\lambda}[{\bar\psi},\psi] = \sum_k \sum_a \left[ i\omega_n - \omega_{\sigma}({\bm k})\right]\,
   {\bar\psi}_{\sigma}(k)\,\psi_{\sigma}(k)\ ,
\label{eq:2.6a}
\ee
with
\be
\omega_{\pm}({\bm k}) = \xi_{\bm k} \mp \lambda\ .
\label{eq:2.6b}
\ee
\ese
Here we see explicitly that the magnetization splits the conduction band into two Stoner bands whose 
Fermi surfaces (FS) are defined by 
\be
\omega_{\sigma}({\bm p})\big\vert_{{\bm p}\,\in\,{\text{FS}_\sigma}} = 0\ ,
\label{eq:2.7}
\ee
and we denote the density of states at the $\sigma$-Fermi surface and the corresponding Fermi
wave number by $\NF^{\sigma}$ and $\kF^{\sigma}$, respectively. In the case of a parabolic band
we have
\bea
\kF^{\pm} &=& \kF\sqrt{1 \pm \lambda/\epsilonF}\ ,
\nonumber\\
\NF^{\pm} &=& \kF^{\pm}\me/2\pi^2\ .
\label{eq:2.7'}
\eea
The Green functions for the two Stoner bands are
\be
G_{\lambda,\sigma}(p) = 1/(i\omega_n - \omega_{\sigma}({\bm p}))\
\label{eq:2.8}
\ee
with $\omega_n = 2\pi T(n+1/2)$ ($n$ integer) a fermionic Matsubara frequency.

The structure of the transverse magnetic susceptibility at small frequencies and wave numbers in an
isotropic ferromagnet is entirely determined by symmetry arguments.\cite{Zinn-Justin_1996} The Goldstone
modes of the spontaneously broken rotational symmetry in spin space are ferromagnons with a
resonance frequency
\be
\omega_0({\bm k}) = D(\lambda)\,{\bm k}^2\ ,
\label{eq:2.9}
\ee
The spin-stiffness coefficient $D$ vanishes as $\lambda\to 0$. It has the dimensions of a diffusion coefficient, 
and is given by a magnetic energy scale divided by a microscopic wave number scale squared, with the latter 
on the order of the Fermi wave number. In the Stoner-Moriya mean-field theory\cite{Moriya_1985} of itinerant 
ferromagnets the former is given by $\lambda$, and for nearly free electrons one obtains 
\bse
\label{eqs:2.9'}
\be
D(\lambda) = \lambda/6\kF^2\qquad (\text{Stoner})\ . 
\label{eq:2.9'a}
\ee
In a Heisenberg spin model with exchange energy $J$ and lattice constant $a$ the corresponding result 
is\cite{Kittel_1996, D_footnote}
\be
D = J\,a^2\qquad (\text{Heisenberg})\ .
\label{eq:2.9'b}
\ee
\ese
If one takes into account mode-mode coupling effects that are not included in the mean-field theory
one finds that $D(\lambda)$ is a nonanalytic function of $\lambda$.\cite{Belitz_et_al_1998, mode_coupling_footnote} The transverse
magnetic susceptibility can be expressed in terms of simple poles that describe circularly polarized ferromagnons,
viz.,
\be
\chi_{\pm}({\bm k},i\Omega) = \frac{K(\lambda)}{(2\NF\Gammat)^2}\,\frac{1}{\omega_0({\bm k}) \pm i\Omega}\ .
\label{eq:2.10}
\ee
The coefficient $K(\lambda)$ is dimensionally an inverse volume. It vanishes as $\lambda\to 0$; in the 
Stoner-Moriya mean-field theory it is given by (see Appendix \ref{app:A})
\bse
\label{eqs:2.10'}
\be
K(\lambda) = 4\NF\lambda \qquad (\text{Stoner})\ . 
\label{eq:2.10'a}
\ee
In a Heisenberg spin model, one has\cite{Forster_1975}
\be
K = 2m\qquad (\text{Heisenberg})\ .
\label{eq:2.10'b}
\ee
\ese
The transverse susceptibility tensor takes the form
\bse
\label{eqs:2.11}
\be
\chi_{\text T}(k) = \frac{1}{2}\left(\begin{array}{cc} \chi_{+}(k) + \chi_{-}(k) & i[\chi_{+}(k) - \chi_{-}(k)] \cr
                                                                              -i[\chi_{+}(k) - \chi_{-}(k)] &  \chi_{+}(k) + \chi_{-}(k) \end{array}\right)
\label{eq:2.11a}
\ee                                                                              
where $k\equiv({\bm k},i\Omega)$. Explicitly one has for small ${\bm k}$ and $\Omega$
\be
\chi_{\text T}(k) = \frac{K(\lambda)}{(2\NF\Gammat)^2}\,\frac{1}{\omega_0({\bm k})^2 - (i\Omega)^2}
   \left( \!\! \begin{array}{cc} D(\lambda){\bm k}^2 & -i(i\Omega) \cr
                                     i(i\Omega) & D(\lambda){\bm k}^2 \end{array} \!\!\! \right).
\label{eq:2.11b}
\ee
\ese
In Appendix \ref{app:A} we show how this structure emerges in an explicit model calculation.

\begin{figure}[t]
\vskip -0mm
\includegraphics[width=6.0cm]{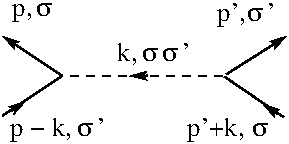}
\caption{Effective electron-electron interaction due to magnon exchange. The dashed line represents
              the effective potential ${\cal V}_{\sigma\sigma'}(k)$.}
\label{fig:1}
\end{figure}

The magnon exchange interaction, Eq.\ (\ref{eq:2.5b}), can now be written
\bse
\label{eqs:2.12}
\be
S_{\text{ex}}[{\bar\psi},\psi] = \frac{1}{2}\sum_{\sigma,\sigma'} \int_k
             \delta n_{\sigma\sigma'}(k)\,{\cal V}_{\sigma'\sigma}(k)\, \delta n_{\sigma'\sigma}(-k)\ .
\label{eq:2.12a}
\ee
Here $\int_k \equiv (1/V)\sum_{\bm k} T\sum_{i\Omega}$, and  the effective potential is given by
\be
{\cal V}_{\sigma \sigma'}(k) = V_{\sigma'\sigma}(k) + V_{\sigma\sigma'}(-k)
\label{eq:2.12b}
\ee
with
\be
V_{\sigma\sigma'}(k) = (1 - \delta_{\sigma\sigma'})\,\Gammat^2\,\chi_{\sigma'}(k)\ ,
\label{eq:2.12c}
\ee
\ese   
This effective interaction is shown diagrammatically in Fig.\ \ref{fig:1}.
Notice that the exchange of magnons couples only
electrons with opposite spin projections, i.e., it leads to inter-Stoner-band scattering only. This is
in contrast to the case of helimagnets, where there is an intra-Stoner-band contribution whose
prefactor is proportional to the square of the helical pitch wave number.\cite{Belitz_Kirkpatrick_Rosch_2006b}

We add a few remarks concerning the validity of this effective action. We have assumed that the conduction
electrons are subject to a magnetization and magnetic fluctuations of unspecified origin whose dynamics are
governed by the physical magnetic susceptibility. Integrating out these
fluctuations leads to an effective action that is purely electronic. Since the feedback of the conduction
electrons on the magnetic susceptibility has already been built into the effective action, the latter must
not be used in ways that constitute, directly or indirectly, a renormalization of the susceptibility; doing
so would constitute double counting. However, it is safe to use the effective action for perturbative
calculations of any observable to first order in the effective potential given by $\Gammat^2\chi$, and we
will use it to calculate the quasiparticle and transport lifetimes to that order. We also note that the
validity of this procedure is more obvious in cases where the magnetization is due to localized electrons
in a band different from the conduction band than in the case of itinerant magnets. However, the coupling
of the spin density to the magnetization fluctuations produced by the other electrons is still the same if
all electrons are in the same band, and with the above caveats the effective action is still valid in that case. 
To illustrate this point we consider the ordinary Fermi-liquid contribution to the electronic relaxation rate, 
as well as the one due to phonons, in Appendix \ref{app:B}, where we
demonstrate that a reasoning for density fluctuations that is analogous to the one given above for 
magnetization fluctuations leads to the standard results for the relaxation rate in these cases.

\subsection{Energy scales}
\label{subsec:II.C}

Before we use the effective action to calculate the single-particle and transport relaxation rates,
let us discuss the relevant energy scales and their relation to experimentally observable quantities.
Here we do so for the simple case of one conduction band; in Sec. \ref{sec:V} we will discuss the more
complicated, and more realistic, situation that arises from the presence of several bands.

The most obvious fundamental magnetic energy scale is the Stoner gap $\lambda$, or the closely
related exchange splitting $\delta E_{\text{ex}} = 2\lambda$.\cite{Stoner_gap_footnote} It can be
measured by photoemission, and also obtained by band structure calculations. The smallest
wave number that can be transferred by means of magnon exchange is $k_0 = \delta E_{\text{ex}}/\vF$.
For a parabolic band this corresponds to $k_0 = \kF^+ - \kF^-$, but the above expression is more
general. The smallest energy that can be transferred by magnon exchange is thus
\be
T_0 = D\,k_0^2 \approx \frac{1}{4}\,D\kF^2 (\delta E_{\text{ex}}/\epsilonF)^2\ .
\label{eq:2.16}
\ee
The largest momentum transfer is given by $k_1 \approx 2\kF$, and we thus have another energy scale,
\be
T_1 = 4D\kF^2\ .
\label{eq:2.17}
\ee
$T_1$ one expects to be close to the exchange splitting; within Stoner theory one has 
$T_1 = 2\lambda/3 = \delta E_{\text{ex}}/3$. Finally, the microscopic energy scale is given by the
Fermi energy $\epsilonF$, and and we have a hierarchy of energy scales, viz., $T_0 \ll T_1 \ll \epsilonF$. 
In particular, the ratio $T_0/T_1$ is given in terms of the Stoner gap in units of the microscopic energy,
\bse
\label{eqs:2.18}
\be
T_0/T_1 \approx \frac{1}{4}\,(\lambda/\epsilonF)^2\ .
\label{eq:2.18a}
\ee
Alternatively, we can use $\ne/2\NF$ as the microscopic energy scale and express the ratio $T_0/T_1$ 
in terms of the magnetization $m$,
\be
T_0/T_1 \approx \frac{1}{9}\,(m/\ne)^2\ .
\label{eq:2.18b}
\ee
\ese
Within Stoner theory this relation holds for $\lambda/\epsilonF \ll 1$, see Eq.\ (\ref{eq:A.7b}), but as
an order-of-magnitude estimate it is expected to hold much more generally.

We finally mention that crystal-field effects break spin-rotational invariance, which gives the
magnons a small gap and leads to yet another energy scale that affects the relaxation rates at
very low temperatures. The magnitude of this effect is highly material dependent, and we 
neglect it for simplicity.

\section{Single-particle relaxation rate}
\label{sec:III}

\begin{figure}[t]
\vskip -0mm
\includegraphics[width=6.0cm]{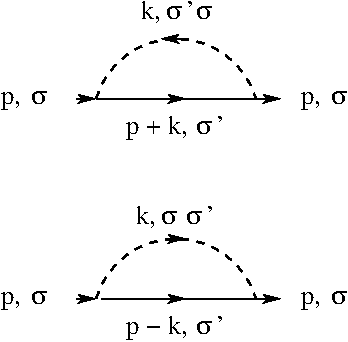}
\caption{Self-energy contributions $\Sigma_{\sigma}(p)$ for the $\sigma$-spin Green function.}
\label{fig:2}
\end{figure}

We now calculate the single-particle inelastic relaxation rate due to the exchange of magnons. To linear
order in the effective potential, Eq.\ (\ref{eq:2.12a}) yields two contributions to the electronic self energy $\Sigma$,
which are shown in Fig. \ \ref{fig:2}.
Analytically, we have
\bea
\Sigma_{\sigma}(p) &=& \int_k \sum_{\sigma'}{\cal V}_{\sigma\sigma'}(k)\,G_{{\lambda},\sigma'}(p+k)
\nonumber\\
                               &=& 2\Gammat^2 \int_k \chi_{\sigma}(k)\,G_{\lambda,-\sigma}(p+k)\ .
\label{eq:3.1}
\eea
Here we have defined the self energy such that the full Green function ${\cal G}$ is given by a
Dyson equation 
\be
{\cal G}_{\sigma}^{-1}(p) = G_{\lambda,\sigma}^{-1}(p) - \Sigma_{\sigma}(p)\ .
\label{eq:3.2}
\ee
Now we consider the single-particle relaxation rate $\Gamma$ for a spin-$\sigma$ quasiparticle, averaged
over the Fermi surface:
\be
\Gamma_{\sigma}(\epsilon) = \frac{-1}{\NF^{\sigma}V} \sum_{\bm p} \delta(\omega_{\sigma}({\bm p}))\,
   \Sigma_{\sigma}''({\bm p},\epsilon)\ ,
\label{eq:3.3}
\ee
where $\Sigma_{\sigma}''({\bm p},\epsilon) = \Im \Sigma({\bm p},i\omega \to \epsilon + i0)$ is the
spectrum of the self energy. Using a spectral representation for the effective potential
and performing the Matsubara frequency sum in Eq.\ (\ref{eq:3.1}) we find
\bse
\label{eqs:3.4}
\bea
\Gamma_{\sigma}(\epsilon) &=& \NF^{-\sigma} \int du\ [n_B(u) + n_F(u + \epsilon)]\,\sum_{\sigma'} {\bar{\cal V}}_{\sigma\sigma'}''(u) 
\nonumber\\
 &=& 2\Gammat^2\,\NF^{-\sigma} \int_{-\infty}^{\infty} du\ [n_B(u) + n_F(u + \epsilon)]\,
   {\bar\chi}_{\sigma}''(u)\ ,
   \nonumber\\
\label{eq:3.4a}
\eea
where $n_B(u) = 1/(e^{u/T} - 1)$ and $n_F(u) = 1/(e^{u/T} + 1)$ are the Bose and Fermi distribution
functions, respectively. Here we have defined
\bea
{\bar{\cal V}}_{\sigma\sigma'}''(u) &=& \frac{1}{\NF^{\sigma}\NF^{\sigma'}V^2} 
                           \sum_{{\bm k},{\bm p}} \delta(\omega_{\sigma}({\bm k}))\,
   \delta(\omega_{\sigma'}({\bm p}))\,
   \nonumber\\
   && \times {\cal V}''_{\sigma\sigma'}({\bm k}-{\bm p},u)
   \label{eq:3.4b}
\eea
and analogously
\be
{\bar\chi}_{\sigma}''(u) = \frac{1}{\NF^{\sigma}\NF^{\sigma'}V^2} \sum_{{\bm k},{\bm p}} \delta(\omega_{\sigma}({\bm k}))\,
   \delta(\omega_{\sigma'}({\bm p}))\,\chi_{\sigma}''({\bm k}-{\bm p},u)\ ,
   \label{eq:3.4c}
\ee
with
\be
\chi_{\pm}''({\bm k},u) = \frac{\mp K(\lambda)\pi}{(2\NF\Gammat)^2}\,\delta(\omega_0({\bm k}) \mp u)
\label{eq:3.4d}
\ee
the spectra of the susceptibilities $\chi_{\pm}$ defined in Eq.\ (\ref{eq:2.10}). We note
the symmetry relation
\be
\NF^+ \,\Gamma_{+}(\epsilon) = \NF^- \, \Gamma_{-}(-\epsilon)\ ,
\label{eq:3.4e}
\ee
\ese
which follows from the symmetry properties of ${\bar\chi}_{\sigma}''(u)$.

Notice that the wave vectors ${\bm k}$ and ${\bm p}$ in Eq.\ (\ref{eq:3.4a}) are pinned to different
Fermi surfaces as a result of the pure inter-Stoner-band scattering mentioned after Eq.\ (\ref{eq:2.12b}).
The spectrum ${\bar\chi}_{\sigma}''(u)$ will therefore be nonzero only for frequencies 
\be
T_0 \leq \vert u \vert \leq T_1\ ,
\label{eq:3.5}
\ee
with $T_0$ and $T_1$ given by Eqs.\ (\ref{eq:2.16}) and (\ref{eq:2.17}).

On the energy shell, 
$\epsilon = 0$, we obtain for the relaxation rate $1/\tau$ on the $\sigma$-Fermi surface
\bea
1/2\tau_{\sigma} &\equiv& \Gamma_{\sigma}(\epsilon=0) = \frac{\pi\,K}{2\NF^{\sigma}T_1}\,T
   \int_{T_0/T}^{T_1/T} \frac{dx}{\sinh x}
\nonumber\\
&& \hskip -40pt = \frac{\pi K}{\NF^{\sigma}T_1} 
     \times \begin{cases} T\,e^{-T_0/T} & \text{if $T\ll T_0$} \cr
                                       \frac{1}{2}\,T\,\ln (T/T_0)    & \text{if $T_0 \ll T \ll T_1$} \cr
                                       \frac{1}{2}\,\ln (T_1/T_0)\,T  & \text{if $T \gg T_1$} \quad.\end{cases}
\nonumber\\
\label{eq:3.6}
\eea
For the thermal resistivity $\rho_{\text{th}} = 1/\kappa$ this implies
\be
\rho_{\text{th}} = \frac{6}{\vF^2 c_V}\,\frac{\pi K/\NF}{T_1} 
   \times \begin{cases} T\,e^{-T_0/T} & \text{if $T\ll T_0$} \cr
                                       \frac{1}{2}\,T\,\ln (T/T_0)    & \text{if $T_0 \ll T \ll T_1$} \cr
                                       \frac{1}{2}\,\ln (T_1/T_0)\,T  & \text{if $T \gg T_1$}\quad . \end{cases}
\label{eq:3.7}
\ee
In Eq.\ (\ref{eq:3.7}) the prefactor is valid in the limit $\lambda \to 0$; more generally there are
corrections of $O((\lambda/\epsilonF)^2)$.
The second line in Eqs.\ (\ref{eq:3.6}) and (\ref{eq:3.7}) is valid to leading logarithmic accuracy only. 
We see that at asymptotically low temperatures the relaxation rate is exponentially small, and that in the
pre-asymptotic temperature window $T_0 \ll T \ll T_1$ there is a logarithmic correction to
the linear behavior. We will further discuss these results in Sec. \ref{sec:V}.

\section{Transport relaxation rate}
\label{sec:IV}

We now turn to the transport relaxation rate, which determines the electrical resistivity. The latter is the
inverse of the electrical conductivity, which is given by the Kubo formula\cite{Mahan_1981}
\bse
\label{eqs:4.1}
\be
\sigma_{ij}(i\Omega)=\frac{i}{i\Omega}[\pi_{ij}(i\Omega)-\pi_{ij}(i\Omega=0)]\ ,
\label{eq:4.1a}
\ee
where the tensor
\bea
\pi_{ij}(i\Omega) &=& -e^2\,T\sum_{n_1,n_2}\frac{1}{V} \sum_{{\bm k},{\bm p}}
v_i({\bm k})\,v_j({\bm p})\,\hskip 0pt
\nonumber\\
&&\hskip -20pt \times \left\langle{\bar\psi}_{n_1,\sigma}({\bm
k})\,\psi_{n_1+n,\sigma}({\bm k})\,{\bar\psi}_{n_2,\sigma'}({\bm
p})\,\psi_{n_2- n,\sigma'}({\bm p})\right\rangle. \nonumber\\
\label{eq:4.1b}
\eea
\ese
is the current-current susceptibility or polarization function. Here ${\bm v}({\bm k}) = \partial\epsilon_{\bm k}/\partial{\bm k}$,
and the average is to be taken with the effective action, Eq.\ (\ref{eq:2.5a}). The four-fermion correlation
function in Eq.\ (\ref{eq:4.1b}) is conveniently expressed in terms of the single-particle Green function
\be
{\cal G}_{\sigma}(p) = 1/(G_{\lambda,\sigma}(p) - \Sigma_{\sigma}(p))
\label{eq:4.2}
\ee
and a vector vertex function ${\bm\Gamma}_{\sigma}$ with components $\Gamma_{\sigma}^i$:
\bea
\pi_{ij}(i\Omega) &=& -ie^2T \sum_{i\omega} \frac{1}{V} \sum_{{\bm p},\sigma} \frac{p^i}{\me} {\cal G}_\sigma({\bm p},i\omega) 
   {\cal G}_\sigma({\bm p},i\omega-i\Omega) 
   \nonumber\\
   && \hskip 70pt \times \Gamma^j_\sigma({\bm p};i\omega,i\omega-i\Omega)\ .
\label{eq:4.3}
\eea
Here we have assumed a quadratic dependence of $\epsilon_{\bm k}$ on ${\bm k}$ for simplicity.
It is important to calculate the vertex function ${\bm\Gamma}$ and the self energy $\Sigma$ in mutually
consistent approximations.\cite{Kadanoff_Baym_1962}  We use the familiar procedure that consists
of a self-consistent Born approximation
for the self energy, which to linear order in the potential $V$ is represented by Eq.\ (\ref{eq:3.1}), and a
ladder approximation for the vertex function,
\bea
{\bm\Gamma}_{\sigma}({\bm p};i\omega,i\omega-i\Omega) &=& i\frac{{\bm p}}{\me} + \frac{T}{V}\sum_{{\bm k},i\Omega^\prime} \sum_{\sigma'} {\cal V}_{\sigma\sigma'}({\bm k}-{\bm p},i\Omega')
\nonumber\\
&&   \hskip -60pt \times  {\cal G}_{\sigma'}({\bm k},i\omega +i\Omega^\prime)\,
                    {\cal G}_{\sigma'}({\bm k},i\omega - i\Omega + i\Omega^\prime) 
   \nonumber \\
&& \hskip -40pt \times {\bm\Gamma}_{\sigma'}({\bm k};i\omega + i\Omega^\prime,i\omega - i\Omega + i\Omega^\prime)\ .
\label{eq:4.4}
\eea
We mention that umklapp processes, which are not explicitly considered here, are necessary in order
to obtain a nonzero transport relaxation rate. In fact, in a Galilean invariant system the electrical resistivity
vanishes due to momentum conservation and the contributions contained in our approximation are cancelled
by terms not included in the ladder approximation. However, the above approximation is effectively valid 
in the presence of umklapp processes, as is the case for Coulomb scattering.\cite{Ziman_1960}
If we define a scalar vertex function $\gamma$ by 
${\bm\Gamma}({\bm p};i\omega,i\omega') = i({\bm p}/\me)\gamma({\bm p};i\omega,i\omega')$, then the Bethe-Salpeter equation
for the latter becomes
\bea
\gamma_{\sigma}({\bm p};i\omega,i\omega-i\Omega) &=& 1 + \frac{T}{V}\sum_{{\bm k},i\Omega^\prime} \sum_{\sigma'}
  \, {\cal V}_{\sigma\sigma'}({\bm p} - {\bm k},i\Omega^\prime)
\nonumber\\
&&\hskip -80pt \times  \frac{{\bm p}\cdot{\bm k}}{{\bm p}^2}\, {\cal G}_{\sigma'}({\bm k},i\omega + i\Omega^\prime)\,
   {\cal G}_{\sigma'}({\bm k},i\omega - i\Omega + i\Omega^\prime)  
\nonumber\\
&& \hskip -40pt \times    \gamma_{\sigma'}({\bm k};i\omega-i\Omega^\prime,i\omega-i\Omega-i\Omega^\prime)\ .
\label{eq:4.5}
\eea

The polarization and conductivity tensors are diagonal, $\sigma_{ij}(i\Omega) = \delta_{ij}\,\sigma(i\Omega)$, and the sum
over Matsubara frequencies in Eq.\ (\ref{eq:4.3}) can be transformed into an integral along the real axis. In the limit of
low temperature, the imaginary part of the self energy, which yields the relaxation rate, goes to zero as we have seen in 
the preceding subsection. The real part just renormalizes the Fermi energy. The relevant limit is thus the one of a
vanishing self energy, and in this limit the leading contributions to the integral come from terms where the frequency
arguments of the two Green functions lie on different sides of the real axis. In the static limit, the Kubo formula for
the conductivity $\sigma = \lim_{\Omega\to 0} \Re\sigma(i\Omega\to\Omega + i0)$, thus becomes
\bea
\sigma &=& \frac{e^2}{3\pi \me^2}\int_{-\infty}^{\infty} \frac{d\epsilon}{4T} \frac{1}{\cosh^2(\epsilon/2T)} \frac{1}{V} \sum_{{\bm p}} {\bm p}^2
\nonumber\\
&&\times  \sum_{\sigma} \vert{\cal G}_{\sigma}({\bm p},\epsilon+i0)\vert^2 \gamma_{\sigma}({\bm p};\epsilon+i0,\epsilon-i0)\ .\qquad
\label{eq:4.6}
\eea
The pole of the Green function ensures that the dominant contribution from the momentum integral comes from the momenta
that obey $\omega_{\sigma}({\bm p}) = \epsilon$. Furthermore, since $\epsilon$ scales as $T$, for the leading $T$-dependence
we can neglect all $\epsilon$-dependencies that do not occur in the form $\epsilon/T$. Equation (\ref{eq:4.6}) then reduces to
\bse
\label{eqs:4.7}
\be
\sigma = \frac{e^2}{2\me}\int_{-\infty}^{\infty} \frac{d\epsilon}{4T} \,\frac{1}{\cosh^2(\epsilon/2T)} 
   \sum_{\sigma} n_{\sigma}\,\frac{\Lambda_{\sigma}(\epsilon)}{\Gamma_{\sigma}(\epsilon)}\ .
\label{eq:4.7a}
\ee
Here $n_{\sigma}$ is the density of the $\sigma$-spin electrons, $\Gamma_{\sigma}$ is the single-particle
rate defined by Eq. (\ref{eq:3.3}), and 
\be
\Lambda_{\sigma}(\epsilon) = \frac{1}{N_F^{\sigma} V}\sum_{{\bm p}} \delta\left(\omega_{\sigma}({\bm p})\right)\,
   \gamma_{\sigma}({\bm p};\epsilon+i0,\epsilon-i0)\ .
\label{eq:4.7b}
\ee
\ese

Using analogous arguments we find, from Eq.\ (\ref{eq:4.5}), that $\Lambda_{\sigma}(\epsilon)$ obeys an
integral equation
\bse
\label{eqs:4.8}
\bea
\Lambda_{\sigma}(\epsilon) &=& 1 + \sum_{\sigma'}\NF^{\sigma'} \int du\ W_{\sigma\sigma'}(u) \left[n_{\text{B}}(u)
   + n_{\text{F}}(u + \epsilon)\right]\,
\nonumber\\
&&\hskip 60pt \times \frac{\Lambda_{\sigma'}(\epsilon+u)}{\Gamma_{\sigma'}(\epsilon+u)}\ ,
\label{eq:4.8a}
\eea
where
\bea
W_{\sigma\sigma'}(u) &=& \frac{1}{\NF^{\sigma} \NF^{\sigma'} V^2} \sum_{{\bm p},{\bm k}} \sum_{\sigma'\neq\sigma} 
   \delta\left(\omega_{\sigma'}({\bm k})\right)\,\delta\left(\omega_{\sigma}({\bm p})\right)\, 
\nonumber\\
&& \hskip 30pt \times   {\cal V}_{\sigma\sigma'}'' ({\bm k} - {\bm p},u) \,{\bm k}\cdot{\bm p}/{\bm p}^2
\label{eq:4.8b}
\eea
\ese
with ${\cal V}_{\sigma\sigma'}''$ the spectrum of the effective potential defined in Eq.\ (\ref{eq:2.12b}).

Now we exploit the fact that ${\bm k}$ and ${\bm p}$ are pinned to the respective Fermi surfaces, and 
use the resulting identity 
$${\bm k}\cdot{\bm p} = \kF^2\,[1 - \omega_0({\bm k}-{\bm p})/2D\kF^2]$$ 
to write
\be
W_{\sigma\sigma'}(u) = (\kF/\kF^{\sigma})^2\,\left[{\bar{\cal V}}''_{\sigma\sigma'}(u) - {\bar{\cal V}}''^{(2)}_{\sigma\sigma'}(u)\right]
\label{eq:4.9}
\ee
with ${\bar{\cal V}}''$ from Eq.\ (\ref{eq:3.4b}), and
\bea
{\bar{\cal V}}''^{(2)}_{\sigma\sigma'}(u) &=& \frac{1}{\NF^{\sigma}\NF^{\sigma'}V^2} 
                           \sum_{{\bm k},{\bm p}} \delta(\omega_{\sigma}({\bm k}))\,
   \delta(\omega_{\sigma'}({\bm p}))\,\frac{\omega_0({\bm k}-{\bm p})}{2D\kF^2}
   \nonumber\\
   && \times {\cal V}''_{\sigma\sigma'}({\bm k}-{\bm p},u)\ .
   \label{eq:4.10}
\eea
Note that the magnon frequency $\omega_0$ in Eq.\ (\ref{eq:4.10}) is equal to $\pm u$ on account of the
spectrum, and therefore ${\bar{\cal V}}''^{(2)}_{\sigma\sigma'}(u)$ has an extra factor of $u$ compared to
${\bar{\cal V}}''(u)$.
${\bar{\cal V}}''$ determines the single-particle rate $\Gamma$
via Eq.\ (\ref{eq:3.4a}), and we define analogously
\be
\Gamma^{(2)}_{\sigma}(\epsilon) = \NF^{-\sigma} \int du\ [n_B(u) + n_F(u + \epsilon)]\,\sum_{\sigma'} {\bar{\cal V}}''^{(2)}_{\sigma\sigma'}(u) 
\label{eq:4.11}
\ee
The integral equation for the vertex function $\Lambda$ now reads
\bea
\Lambda_{\sigma}(\epsilon) &=& 1 + \left(\frac{\kF}{\kF^{\sigma}}\right)^2 \int du \sum_{\sigma'} \NF^{\sigma'}\left[{\bar{\cal V}}''_{\sigma\sigma'}(u)
   - {\bar{\cal V}}''^{(2)}_{\sigma\sigma'}(u)\right]
   \nonumber\\
&& \times \left[n_{\text{B}}(u) + n_{\text{F}}(u + \epsilon)\right]\,\frac{\Lambda_{\sigma'}(u + \epsilon)}{\Gamma_{\sigma'}(u + \epsilon)}
   \ ,
\label{eq:4.12}   
\eea

For the case of a spin-independent potential, Eq.\ (\ref{eq:4.12}) reduces to the integral equation familiar from the 
electron-phonon scattering problem; only the $u$-dependence of the kernel is different. This integral equation is 
usually solved in the seemingly uncontrolled approximation that replaces
$\Lambda(u+\epsilon)/\Gamma(u+\epsilon)$ on the right-hand side by $\Lambda(\epsilon)/\Gamma(\epsilon)$, turning the
integral equation into an algebraic equation. In Ref.\ \onlinecite{Belitz_Kirkpatrick_2010a} two of the present authors have
shown that the integral equation can be solved asymptotically exactly, that the exact solution yields a result for the
conductivity that coincides with the lowest-order variational solution of the Boltzmann equation, and that the simple
approximation yields the same low-temperature dependence (albeit with a different prefactor) as the exact solution. The proof
of these statements can be generalized to the current case of a two-by-two matrix equation. For the purpose of deriving
the low-temperature behavior, we thus can employ the approximation, which turns Eq.\ (\ref{eq:4.12}) into two coupled
algebraic equations for $\Lambda_{\pm}(\epsilon)$. Since the prefactor of the temperature dependence of the 
conductivity is approximation-dependent anyway, we can put $\epsilon = 0$ and use the temperature-dependent
rates $\Gamma_{\sigma} \equiv \Gamma_{\sigma}(\epsilon=0)$ and vertices $\Lambda_{\sigma} \equiv
\Lambda_{\sigma}(\epsilon=0)$ in the Kubo formula, Eq.\ (\ref{eq:4.7a}). $\Lambda_{\sigma}$ then obeys
\bea
\Lambda_{+} &=& 1 + (\kF/\kF^+)^2 \left[\Gamma_{+} - \Gamma^{(2)}_{+}\right]\,
                                  \Lambda_{-}/\Gamma_{-}\ ,
\nonumber\\                                 
\Lambda_{-} &=& 1 + (\kF/\kF^-)^2 \left[\Gamma_{-} - \Gamma^{(2)}_{-}\right]\,
                                  \Lambda_{+}/\Gamma_{+}\ .
                                  \nonumber\\
\label{eq:4.13}
\eea
$\Gamma_{\sigma}$ is given by Eq.\ (\ref{eq:3.6}), and $\Gamma_{\sigma}^{(2)}$ is given by an analogous
integral with an additional power of the frequency in the integrand. We find
\bea
\Gamma_{\sigma}^{(2)} &=& \frac{\pi K}{\NF^{\sigma}}\,\frac{T^2}{T_1^2}
   \int_{T_0/T}^{T_1/T} dx\,\frac{x}{\sinh x}
\nonumber\\
&& \hskip -40pt = \frac{\pi K}{\NF^{\sigma}}\,\frac{T}{T_1} 
     \times \begin{cases} \frac{2T_0}{T_1}\left(1 + \frac{T}{T_0}\right)\,e^{-T_0/T} & \text{if $T\ll T_0$} \cr
                                       \frac{\pi^2}{4}\,T/T_1    & \text{if $T_0 \ll T \ll T_1$} \cr
                                       1  & \text{if $T \gg T_1\quad.$} \end{cases} 
\nonumber\\
\label{eq:4.14}
\eea
Comparing with Eq.\ (\ref{eq:3.6}) we see that for asymptotically small $T$, $\Gamma^{(2)}$ is proportional
to $\Gamma$ with a small factor of proportionality $2T_0/T_1 \ll 1$, whereas for
$T_0 \ll T \ll T_1$ it carries an additional factor of temperature. 

We now solve the equations (\ref{eq:4.13}). Neglecting $\lambda/\epsilonF \ll 1$ wherever it is not of qualitative
importance, we find
\be
\frac{\Lambda_{\pm}}{\Gamma_{\pm}} = \frac{\Gamma_{+} + \Gamma_{-} - \Gamma^{(2)}_{\pm}}
   {-\frac{4T_0}{T_1}\,\Gamma_{+}\Gamma_{-} + \Gamma_{+}\Gamma_{-}^{(2)} + \Gamma_{-}\Gamma_{+}^{(2)}
      - \Gamma_{+}^{(2)}\Gamma_{-}^{(2)}}\ .
\label{eq:4.15}
\ee
Equation (\ref{eq:3.4e}) allows us to express $\Lambda_{\sigma}/\Gamma_{\sigma}$ entirely in terms of 
$\Gamma_{\sigma}$ and $\Gamma^{(2)}_{\sigma}$, Neglecting all prefactors that just give small corrections
of $O(\lambda/\epsilonF)$ to factors of $O(1)$ we finally obtain a transport relaxation time
\bse
\label{eqs:4.16}
\be
\tau_{\text{tr}} = \frac{\Gamma - \Gamma^{(2)}/2}{-\frac{4T_0}{T_1}\,(\Gamma)^2 + 2\Gamma\Gamma^{(2)} 
      - (\Gamma^{(2)})^2}\ ,
\label{eq:4.16a}
\ee
in terms of which the electrical conductivity is given by a Drude formula
\be
\sigma = \frac{\ne e^2}{\me}\,\tau_{\text{tr}}\ ,
\label{eq:4.16b}
\ee
\ese

Here $\Gamma \approx \Gamma_{+} \approx \Gamma_{-}$ and 
$\Gamma^{(2)} \approx \Gamma^{(2)}_{+} \approx \Gamma^{(2)}_{-}$ are given by Eq.\ (\ref{eq:3.6})
and Eq.\ (\ref{eq:4.14}), respectively, with $\NF^{\sigma}$ replaced by  $\NF$. Note that our approximations
have affected overall prefactors only, but not the relative prefactors of the four terms in the denominator
in Eq.\ (\ref{eq:4.16a}). Comparing Eq.\ (\ref{eq:3.6}) with (\ref{eq:4.14}) we see that, for $T\ll T_0$,
$\Gamma^{(2)}$ is proportional to $\Gamma$:
\be
\Gamma^{(2)} = \frac{2T_0}{T_1}\,(1 + T/T_0)\,\Gamma\ ,
\label{eq:4.17}
\ee
and an inspection reveals that the leading contributions among the three terms in the denominator in
Eq.\ (\ref{eq:4.16a}) cancel, which leads to $1/\tau_{\text{tr}} \propto T^2\,\exp(-T_0/T)$. At asymptotically
low temperatures $\tau_{\text{tr}}$ is therefore {\em not} proportional to $1/\Gamma^{(2)}$, but rather
carries an extra factor of $T_0/T$. For the contribution to the electrical resistivity $\rho_{\text{el}} = 1/\sigma$ due to
magnon exchange we finally obtain
\begin{widetext}
\be
\rho_{\text{el}}(T) = \frac{\me}{\ne\,e^2}\,\frac{\pi K}{\NF T_1} \times
   \begin{cases} (4/T_1)\,T^2\, e^{-T_0/T} & \text{if $T \ll T_0$} \cr
                          (\pi^2/2T_1)\,T^2 & \text{if $T_0 \ll T \ll T_1$} \cr
                          T & \text{if $T \gg T_1$}\quad.
  \end{cases}
\label{eq:4.18}
\ee  
\end{widetext}   
We see that in the preasymptotic temperature window $T_0 \ll T \ll T_1$ we recover the $T^2$ behavior that was obtained
in Ref.\ \onlinecite{Ueda_Moriya_1975}, but for asymptotically low temperatures we obtain an exponentially small result   
that has the form of Eq.\ (\ref{eq:1.1}). We will discuss this result in the next section.

\section{Discussion}
\label{sec:V}

To summarize our results, we have presented a very general theory of electron relaxation due to the exchange
of magnons in metallic ferromagnets. The theory is valid for both itinerant ferromagnets, where the magnetization
is due to the conduction electrons themselves, and for localized-moment ferromagnets, where the magnetization is due to
localized spins in a different band. We have found that at asymptotically low temperatures, below a temperature
scale $T_0$, both the single-particle relaxation rate and the transport relaxation rate are exponentially small.
This behavior carries over to the magnon-exchange contributions to the thermal and electrical resistivities, 
which are determined by these respective
rates. The exponential temperature dependence is a direct consequence of the split conduction band in a
metallic ferromagnet. In a preasymptotic temperature regime $T_0 \ll T \ll T_1$, with $T_1$ close to the 
exchange splitting, we recover the $T^2$ behavior of the transport rate that was found in 
Ref.\ \onlinecite{Ueda_Moriya_1975}. The single-particle rate is proportional to $T$ in this regime. For 
$T\gg T_1$ the two rates both show a linear temperature dependence. 

We start our discussion of these results by recalling the physical reason for the exponential dependence
at low temperatures. 
\begin{figure}[b]
\vskip -0mm
\includegraphics[width=8.5cm]{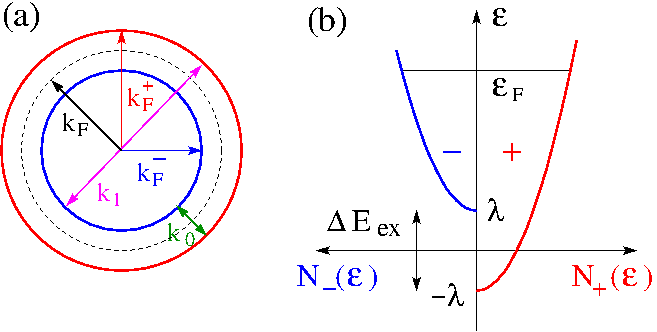}
\caption{Fermi surfaces and associates Fermi wave numbers (a) and densities of states (b) for the up- and 
down-spin electrons. $\lambda$ is the Stoner gap, and $\Delta E_{\text{ex}}$ is the exchange splitting. $k_0$
and $k_1$ are the smallest and largest transferrable wave numbers, respectively. See the text for more
explanation.}
\label{fig:3}
\end{figure} 
Figure \ref{fig:3} schematically
shows the split conduction band (a), and the densities of states for the up (+) and down (-) spin electrons (b),
for the case of a spherical Fermi surface.
Since the magnons couple only electrons with opposite spin, the smallest transferrable wave number is
$k_0 =\kF^{+} - \kF^{-} \approx \Delta E_{\text{ex}}/\vF$. Given the magnon dispersion relation, $\omega = D k^2$,
this translates into a smallest transferrable energy $T_0 = D\,k_0^2$, and since the magnon stiffness
coefficient $D$ is itself roughly proportional to $\Delta E_{\text{ex}}$, we have $T_0 \propto (\Delta E_{\text{ex}})^3$.
For temperatures $T \ll T_0$ the relaxation rates will thus show activated behavior with an activation energy $T_0$.
The exponential behavior is multiplied by a power law that cannot be captured by elementary arguments. The
largest momentum transfer is given by $k_1 = \kF^{+} + \kF^{-} \approx 2\kF$, and the corresponding largest
energy transfer is $T_1 = D\,k_1^2 \approx \Delta E_{\text{ex}}$. $T_1$ is the fundamental magnetic energy
scale, analogous to the Debye temperature $\Theta_D$ in the case of electron-phonon coupling. $T_0$ has no analog in 
the electron-phonon problem. For $T \ll T_1$ the transport-relaxation rate is small compared to the single-particle
rate by a factor of $T/T_1$. This is analogous to the electron-phonon case, where the corresponding factor is
$(T/\Theta_D)^2$. The difference between our results and those of Ueda and Moriya, Ref.\ \onlinecite{Ueda_Moriya_1975},
can be traced to the fact that these authors neglected the exchange splitting in the final stages of their
calculation. As a result, they obtained a $T^2$ behavior of the transport relaxation rate at low temperatures, which in fact is valid
only for temperatures larger than $T_0$. Note that this discrepancy pertains to the magnon or spin-wave contribution
to the electrical resistivity only. The contributions from dissipative excitations, which we have not discussed, have been found to be
unaffected by the exchange splitting and proportional to $T^2$ even at asymptotically low temperatures.\cite{Ueda_Moriya_1975}

For the power-law behavior at $T \gg T_0$ the quadratic spectrum of the magnons is important, and also the coupling of the
electrons to the magnetic fluctuations. Comparing with the case of helical magnets,\cite{Belitz_Kirkpatrick_Rosch_2006b}
we notice one important difference with respect to the latter. In either case the Goldstone mode is a phase fluctuation, but
in the ferromagnon case the electron spin density couples directly to the phase, whereas in the helimagnon case the
coupling is to the gradient of the phase. This is because in the helimagnon case the dominant low-$T$ contribution to
the scattering rates comes from intra-Stoner-band scattering. Within a given band, the phase itself has no physical meaning, 
and the coupling therefore involves a gradient. In the ferromagnetic case, on the other hand, we deal with inter-Stoner-band
scattering. The coupling therefore effectively is to the difference of two phases, which does have a physical meaning. We 
note in passing that this latter notion also manifests itself in a spin Josephson effect, 
see Ref.\ \onlinecite{Nogueira_Bennemann_2004}.

We now turn to estimates of the values of $T_1$ and $T_0$. To get an idea about the order of magnitude of these
temperature scales, let us first consider the fictitious case of simple (i.e., single-conduction-band) metals with magnetic
properties as in the classic ``high-temperature'' ferromagnets nickel, cobalt, and iron. The values of the exchange 
splitting in these materials, as determined by photoemission,
are $\Delta E_{\text{ex}} \approx 0.25$ eV, $1.0$ eV, and $2.0$ eV, respectively.\cite{Himpsel_1991, Himpsel_et_al_1998}
Values for the spin-stiffness coefficient $D$ in meV$\AA^2$ obtained from neutron scattering are 
$364$ for Ni, $500$ for Co, and $281$ for Fe.\cite{Kittel_1996} With a generic value $\kF \approx 1\AA^{-1}$ for the Fermi wave
number, and $\epsilonF \approx 10^5$ K for the Fermi energy, Eqs.\ (\ref{eq:2.16}) and (\ref{eq:2.17}) yield
$T_1 \approx 10,000 - 20,000$ K for these materials, and $T_0 \approx 500$ mK for Ni, $10$ K for Co, and
$30$ K for Fe. Estimates of the ratio $T_0/T_1$ using the relation (\ref{eq:2.18b}) yields similar results. Notice that the prefactor
$\pi K/\NF T_1$ in Eq.\ (\ref{eq:4.18}) is of order unity, so the prefactor of the $T^2$ behavior of the resistivity is larger
than the Fermi-liquid $T^2$ contribution (see Eq.\ (\ref{eq:B.7})) by roughly a factor of $\epsilonF/T_1 \approx 10$ in a 
single-band  model.

Also of interest are weak ferromagnets, such as MnSi,\cite{MnSi_footnote} or Ni$_3$Al, where 
$D \approx 23.5$ meV$\AA^2$ (MnSi)\cite{Boni_Roessli_Hradil_2011} and $D \approx 70$ meV$\AA^2$
(Ni$_3$Al),\cite{Bernhoeft_et_al_1982} respectively. The magnetic moments, $0.4\mu_{\text{B}}$ per
formula unit for MnSi,\cite{Pfleiderer_2007} and $0.17\mu_{\text{B}}$ for Ni$_3$Al,\cite{Bernhoeft_et_al_1982} are about 
two thirds and one third, respectively, of that of Ni. Given the observed near-linear correlation between the 
magnetic moment and the exchange splitting,\cite{Himpsel_1991} this suggests $\Delta E_{\text{ex}} \approx 0.17$ eV
for MnSi,\cite{MnSi_splitting_footnote}, and $\Delta E_{\text{ex}} \approx 0.07$ eV for Ni$_3$Al. If
we use again $\kF \approx 1\AA^{-1}$ and $\epsilonF \approx 10^5$ K, this yields $T_1 \approx 1,000$ K and
$T_0 \approx 20$ mK for MnSi, and $T_1 \approx 2,800$ K and $T_0 \approx 10$ mK for Ni$_3$Al.

In reality, all of these materials are transition metals, or compounds containing transition metals, 
with a complicated band structure and Fermi surfaces that consist of multiple sheets. One consequence of this is that 
the electron-electron scattering contribution to the electrical resistivity is likely much larger than a single-band model 
would imply, and it has been suggested that it makes the largest contribution to the observed $T^2$ behavior at low 
temperatures.\cite{Campbell_Fert_1982} The reason is that different band
edges have different distances from the common chemical potential, which in effect leads to different Fermi
temperatures. Depending on whether or not the various scattering processes flip the electron spin, and whether
or not they couple different sheets of the Fermi surface, the relaxation rates or the relaxation times may be
additive, which leads a complicated structure of the overall resistivity. In addition, there are the contributions from
the dissipative spin excitations, which also are proportional to $T^2$.\cite{Ueda_Moriya_1975} As a result, the low-temperature
transport rate in Fe, Ni, and Co is about 100 times larger than one would expect from the Coulomb contribution in a single-band
model with a single Fermi temperature of about $10^5$K.\cite{Campbell_Fert_1982}
$T_1$, on the other hand, is largely unaffected by a complicated band structure: It is given by $D$ times the largest
possible momentum transfer squared, see Sec.\ \ref{subsec:II.C}, and in a good metal the latter is on the order of
$2\pi/a$, with $a$ the lattice constant, which is close to the value of $2\kF$ for a single spherical Fermi surface
that yields the same electron density. The estimates of the temperature scale $T_1$, which is the magnetic analog
of the Debye temperature for phonons, given above are therefore model independent and depend only
on the experimentally measured spin stiffness coefficient.

As a result, we expect the magnon contribution to the electrical resistivity in Fe, Ni, and Co at temperatures 
$T > T_0$ to be about an order of magnitude {\em less} than the combined contribution from the Coulomb
interaction and the dissipative magnetic excitations. 
In MnSi and Ni$_3$Al $T_1$ is much lower and the magnon scattering is accordingly stronger. However,
the observed prefactors of the $T^2$ term in the resistivity of MnSi and Ni$_3$Al are orders of magnitude
larger than even the ones in Fe, Ni, and Co, and the same is true for the weak ferromagnet
ZrZn$_2$.\cite{Pfleiderer_et_al_1997, Ogawa_1976} The prefactor $\pi K/\NF T_1$ in Eq.\ (\ref{eq:4.18})
is expected to be of $O(1)$ not just in model calculations, but also in real materials, since both $K$ and
$T_1$ correlated roughly with the magnetization. Given the above discussion of the relatively narrow
range of plausible values of $T_1$, we conclude that the experimental value of the prefactor of the $T^2$
term in the electrical resistivity of weak ferromagnets cannot possibly be explained by electron-magnon scattering.
We emphasize again, however, that these considerations do not take into account the scattering of electrons
by dissipative magnetization fluctuations, which lead to a $T^2$ contribution to the resistivity even at low
temperatures and whose prefactor is not as universal as that of the magnon-exchange contribution. A
corresponding statement holds for the Coulomb contribution.

For $T_0$ the influence of the band structure is more complicated. Consider the effective potential given by
Eqs.\ (\ref{eq:3.4b} - \ref{eq:3.4d}). If the up-spin and down-spin electrons, respectively, belong to different
bands with different effective masses, then there will be a lower cutoff for the frequency $u$ even in the
limit of a vanishing Stoner gap, $\lambda \to 0$. For magnon-exchange scattering between electrons in
Stoner-subbands of the same band, on the other hand, the structure of the calculations in Secs.\ \ref{sec:III}
and \ref{sec:IV} is unchanged. We therefore expect different values of $T_0$ for the various scattering
processes that involve electrons on different sheets of the Fermi surface.

The following picture now emerges. With decreasing temperature, contributions to the magnon-exchange
part of the electronic scattering rate will sequentially freeze out as the temperature drops below a sequence
of temperature scales $T_0$. Rough estimates for the lowest of these temperature scales have been given
above; estimating the higher ones requires a detailed analysis of the band structure. Below this lowest $T_0$
the magnon contribution to both the transport rate and the single-particle rate will be exponentially small, leaving
the Coulomb contribution and the one from dissipative magnetization fluctuations as the most obvious 
candidates for a $T^2$ behavior. Experimentally, this is expected
to manifest itself in a distinct temperature dependence of the prefactor of the $T^2$ term in the electrical
resistivity. It is desirable for the relevant temperature scales to be small enough that phonon contributions are
negligible. In that respect, Fe, Ni, and Co are not ideal. In MnSi, the helical nature of the magnetic phase is
expected to manifest itself on the temperature scale given by $T_0$. This leaves Ni$_3$Al, or other true
weak ferromagnets as the most promising candidates for observing this consequence
of the exchange splitting in a metallic ferromagnet. We stress, however, that according to the above
discussion the magnon contribution to the electrical resistivity in weak ferromagnets is likely dwarfed by 
other contributions. 

Another possible effect of a complicated band structure is that there may be points or lines in
reciprocal space where the two Stoner band cross. This will weaken the exponential suppression of the
relaxation rates, but the weakening will depend on the nature of the crossing.

We finally mention that the interplay of quenched disorder with the scattering
processes discussed above constitutes an interesting problem that is likely important for a quantitative
understanding of real materials. For fairly strong disorder, $\lambda\tau_{\text{el}}\ll 1$ with $\tau_{\text{el}}$
the elastic scattering time, the theory of Ref.\ \onlinecite{Kirkpatrick_Belitz_2000} applies and it is easy to see
that there is no exponential suppression of the magnon contribution to the relaxation rates at low temperature.
A complete discussion of disorder effects constitutes a separate problem.

\acknowledgments
We thank Achim Rosch for discussions. This work was supported by the NSF under grants number DMR-09-01952 and
DMR-09-01907. Part of this work was performed at the Aspen Center for Physics and supported by the NSF under grant
number PHYS-1066293.

\appendix

\section{Stoner model for itinerant ferromagnets}
\label{app:A}

In this appendix we show how to recover the Stoner-Moriya results\cite{Moriya_1985} for itinerant
ferromagnets within the present formalism. Our starting point is a fermionic action
\be
S[{\bar\psi},\psi] = S_0[{\bar\psi},\psi] +
\frac{\Gamma_{\text{t}}}{2}\int dx\ {\bm n}_{\text{s}}(x)\cdot{\bm n}_{\text{s}}(x)\ ,
\label{eq:A.1}
\ee
with $\Gammat$ the spin-triplet interaction amplitude that is responsible for ferromagnetism. Our notation is
the same as in Sec.\ \ref{sec:II}.	

\subsection{Mean-field approximation}
\label{subsec:A.1}

A simple mean-field approximation that describes ferromagnetic order, and its coupling to the electron
spin density, consists of replacing one of the spin density fields in Eq.\ (\ref{eq:A.1}) by its expectation
value according to 
\be
{\bm n}_{\text{s}}^2 \approx 2\langle{\bm n}_{\text{s}}\rangle\cdot{\bm n}_{\text{s}} - \langle{\bm n}_{\text{s}}\rangle^2\ .
\label{eq:A.2}
\ee
If we take the magnetic order to be in the 3-direction, $\langle n_{\text{s},i}\rangle = \delta_{i3}\,\lambda/\Gamma_{\text{t}}$, 
then this approximation amounts to replacing the action $S$ with an action $S_{\lambda}$ that describes electrons
with no spin-triplet interaction subject to a magnetic field of strength $\lambda$ in the 3-direction:
\be
S_{\lambda}[{\bar\psi},\psi] = S_0[{\bar\psi},\psi] + \lambda \int dx\ n_{\text{s},3}(x)\ .
\label{eq:A.3}
\ee
The remaining question pertains to the action one should use to calculate $\langle{\bm n}_{\text{s}}\rangle$; this
choice determines $\lambda$. The usual self-consistent mean-field requirement stipulates that this
average be determined by $S_{\lambda}$ itself:
\bea
\lambda &=& \Gamma_{\text{t}} \langle n_{\text{s},3}(x)\rangle_{\lambda} 
                      \equiv \frac{\Gamma_{\text{t}}}{Z_{\lambda}}\int D[{\bar\psi},\psi]\ n_{\text{s},3}(x)\ e^{S_{\lambda}[{\bar\psi},\psi]}
\nonumber\\
              &=& \Gamma_{\text{t}} \frac{d}{d\lambda}\,\log Z_{\lambda}\ .
\label{eq:A.4}
\eea
For simplicity, we take $S_0$ to describe
free electrons. That is, we neglect all electron-electron interactions that are not crucial for
magnetism, and we assume a parabolic band; a generalization to band electrons is 
straightforward. The Green function corresponding to $S_0$ then is
\be
G_0({\bm k}, i\omega_n) = \frac{1}{i\omega_n-\xi_{\bm k}}\ ,
\label{eq:A.5}
\ee
and the self-consistency condition, Eq. (\ref{eq:A.4}), takes the form
\be
1 = -2\Gamma_{\text{t}} \int_p \frac{1}{G_0^{-2}(p) - \lambda^2}\ .
\label{eq:A.6}
\ee
We recognize this as the equation of state of Stoner theory, with $\lambda$ the Stoner gap.
The condition for a nonzero solution for $\lambda$ is $2\NF\Gamma_{\text{t}} > 1$, and by performing the integral we
find explicitly
\bse
\label{eqs:A.7}
\be
\lambda = 2\NF\Gammat\,\frac{\epsilonF}{3}\,\left[(1+\lambda/\epsilonF)^{3/2} - (1-\lambda/\epsilonF)^{3/2}\right]\ .
\label{eq:A.7a}
\ee
If we recall that the magnetization is given as $m = \lambda/\Gammat$ (see Eqs.\ (\ref{eq:2.1'}, \ref{eq:2.2a})), we
can write this result as
\be
m = 2\NF\lambda\,\left[1 + O((\lambda/\epsilonF)^2\right]\ .
\label{eq:A.7b}
\ee
\ese

The action $S_{\lambda}$ contains information about the long-range order, but does not contain
any ferromagnetic fluctuations. It will serve as a building block for the effective action, and we will
refer to it as the ``reference ensemble''.

We now determine the spin susceptibility 
\be
\chi^{}_{\lambda,ij}(x,y) = \left\langle \delta n_{\text{s},i}(x)\,\delta n_{\text{s},j}(y) \right\rangle_{S_{\lambda}}\ ,
\label{eq:A.8}
\ee
associated with the reference ensemble. In terms of the reference-ensemble Green function
\be
G_{\lambda}(k) = \frac{G_0^{-1}(k)}{G_0^{-2}(k) - \lambda^2}\,\sigma_0 - \frac{\lambda}{G_0^{-2}(k) - \lambda^2}\,\sigma_3\ ,
\label{eq:A.9}
\ee
$\chi_{\lambda}$ can be written
\be
\chi_{\lambda,ij}(x,y) = - \tr\left[\sigma_i\, G_{\lambda}(x,y)\, \sigma_j\, G_{\lambda}(y,x)\right]
\label{eq:A.10}
\ee
where the trace is over the spin degrees of freedom. Evaluating the trace, and performing a Fourier
transform, we find
\be
\chi^{}_{\lambda,ij}(k) = \left( \begin{array}{ccc}
                          \ \ f_1(k) & \ \ f_2(k) & 0 \\
                          - f_2(k) & \ \ f_1(k) & 0 \\
                          0                                    &     0                                     &f_3(k)
                          \end{array}\right)\ ,
\label{eq:A.11}
\ee
where
\bse
\label{eqs:A.12}
\bea
f_1(k) &=& -2\int_p 
   \frac{G^{-1}_0(p)G^{-1}_0(p-k) - \lambda^2}
       {[G^{-2}_0(p) - \lambda^2][G^{-2}_0(p-k) - \lambda^2]}\ ,
       \nonumber\\
\label{eq:A.12a}\\
f_2(k) &=& -2i\lambda\int_p 
    \frac{G^{-1}_0(p) - G^{-1}_0(p-k)} {[G^{-2}_0(p) - \lambda^2][G^{-2}_0(p-k) - \lambda^2]}\ ,
    \nonumber\\
\label{eq:A.12b}\\
f_3(k) &=& - 2 \int_p
   \frac{G^{-1}_0(p) G^{-1}_0(p-k) + \lambda^2} {[G^{-2}_0(p) - \lambda^2][G^{-2}_0(p-k) - \lambda^2]}\ .   
   \nonumber\\
\label{eq:A.12c}
\eea
\ese
We note that 
\bse
\label{eqs:A.13}
\bea
f_1(k=0) &=& 1/\Gamma_{\text{t}}\ ,
\label{eq:A.13a}\\
f_2(k=0) &=& 0\ ,
\label{eq:A.13b}
\eea
\ese
with the first equality following from the equation of state, Eq.\ (\ref{eq:A.6}).

\subsection{Physical spin susceptibility, and Goldstone modes}
\label{subsec:A.2}

The reference ensemble does not reflect the magnons that are the Goldstone modes of
the spontaneously broken symmetry in the ferromagnetic phase. To describe the magnons
we need a theory of fluctuations that is consistent with the treatment of the static magnetization. 
Quite generally, a Gaussian approximation for the order-parameter fluctuations is consistent with
a mean-field treatment of the order parameter itself.\cite{Landau_Lifshitz_V_1980}
To determine the former we first note that the reference ensemble spin susceptibility $\chi_{\lambda}$
corresponds to a Gaussian fluctuation action
\bse
\label{eqs:A.14}
\be
{\cal A}_{\lambda,\text{fluct}}[\delta {\bm n}_{\text{s}}] = \frac{-1}{2} \int dx dy\ \delta n_{\text{s},i}(x)\,\chi_{\lambda,ij}^{-1}(x,y)\,\delta n_{\text{s},j}(y)
\label{eq:A.14a}
\ee
that generates $\chi^{}_{\lambda}$ via
\be
\chi_{\lambda,ij}^{}(x,y) = \int D[\delta{\bm n}_{\text{s}}]\ \delta n_{\text{s},i}(x)\,\delta n_{\text{s},j}(y)\ e^{-{\cal A}_{\lambda,\text{fluct}}[\delta{\bm n}_{\text{s}}]}\ .
\label{eq:A.14b}
\ee
\ese
To this we need to add the fluctuation contribution from the original spin-triplet interaction in Eq.\ (\ref{eq:A.1}).
The Gaussian fluctuation action then reads
\bse
\label{eqs:A.15}
\be
{\cal A}_{\text{fluct}}[\delta {\bm n}_{\text{s}}]  = \frac{-1}{2} \int dx dy\ \delta n_{\text{s},i}(x)\,\chi^{-1}_{ij}(x,y)\,\delta n_{\text{s},j}(y)
\label{eq:A.15a}
\ee
with the physical spin susceptibility $\chi$ given by
\be
\chi^{-1}_{ij}(x,y) = \chi_{\lambda,ij}^{-1}(x,y) - \delta_{ij}\,\Gamma_{\text{t}}\ .
\label{eq:A.15b}
\ee
\ese
Focusing on the transverse (T) channel ($i=1,2$), and performing a Fourier transform, we have
\begin{widetext}
\bse
\label{eqs:A.16}
\be
\chi^{-1}_{\text{T}}({\bm k},i\Omega_n) = \left(\begin{array}{cc} 
   f_1({\bm k},i\Omega_n)/N({\bm k},i\Omega_n) - \Gamma_{\text{t}} & -f_2({\bm k},i\Omega_n)/N({\bm k},i\Omega_n) \\
   f_2({\bm k},i\Omega_n)/N({\bm k},i\Omega_n) & f_1({\bm k},i\Omega_n)/N({\bm k},i\Omega_n) - \Gamma_{\text{t}}
   \end{array}\right)\ .
\label{eq:A.16a}
\ee
where
\be
N({\bm k},i\Omega_n) = \left(f_1({\bm k},i\Omega_n)\right)^2 + \left(f_2({\bm k},i\Omega_n)\right)^2\ .
\label{eq:A.16b}
\ee
\ese
From Eqs.\ (\ref{eqs:A.13}) we see that $\chi^{-1}_{\text{T}}$ at zero frequency and wave number has two zero eigenvalues.
These reflect the two Goldstone modes. Expanding to linear order in $i\Omega$ and to second order in ${\bm k}$
we find explicitly
\bse
\label{eqs:A.17}
\be
\chi^{-1}_{\text{T}}({\bm k},i\Omega_n) = \frac{(2\NF\Gamma_{\text{t}})^2 }{2\NF}\,
                                         \left(\begin{array}{cc} {\hat{\bm k}}^2 f_{\bm k}(\lambda)/3 &  2i(i{\hat\Omega}_n) f_{\Omega}(\lambda)\epsilonF/\lambda \\
                                                                           -2i(i{\hat\Omega}_n) f_{\Omega}(\lambda) \epsilonF/\lambda    & {\hat{\bm k}}^2 f_{\bm k}(\lambda)/3\end{array}\right),
\label{eq:A.17a}
\ee
where ${\hat{\bm k}} = {\bm k}/2\kF$, ${\hat\Omega} = \Omega/4\epsilonF$, and
\bea
f_{\bm k}(\lambda) &=& \frac{-4\epsilonF^3}{5\lambda^3}\,
   \left[\left(1 - \frac{3\lambda}{2\epsilonF}\right)\left(1 +  \frac{\lambda}{\epsilonF}\right)^{3/2} 
   - \left(1 + \frac{3\lambda}{2\epsilonF}\right)\left(1 -  \frac{\lambda}{\epsilonF}\right)^{3/2}\right],
\label{eq:A.17b}\\
f_{\Omega}(\lambda) &=& \frac{\epsilonF}{3\lambda}\,\left[\left(1 + \frac{\lambda}{\epsilonF}\right) ^{3/2}
   - \left(1 - \frac{\lambda}{\epsilonF}\right)^{3/2}\right]\ .
   \nonumber\\
\label{eq:A.17c}
\eea
\ese
\end{widetext}
Physically, the Stoner gap is always small compared to the Fermi energy, and it therefore is useful to consider
the limit of weak ferromagnets, $2\NF\Gamma_{\text{t}} \approx 1$ and $\lambda/\epsilonF \ll 1$, where we have
\be
f_{\bm k}(\lambda \to 0) = f_{\Omega}(\lambda \to 0) = 1 + O(\lambda^2)\ .
\label{eq:A.18}
\ee
Inverting Eq.\ (\ref{eq:A.17a}) we obtain the transverse physical spin susceptibility in the form given in Eq.\ (\ref{eq:2.11b}),
with the mean-field values for $K(\lambda)$ and $D(\lambda)$ as quoted in the main text. We note that the
spin precession effect, which is represented by the off-diagonal matrix elements in Eq.\ (\ref{eq:A.17a}) and leads
to the characteristic $\omega \propto {\bm k}^2$ dispersion relation of ferromagnetic magnons, appears in a
rather elementary way in this treatment of itinerant electrons. In spin models, by contrast, it emerges from
a topological contribution to the action.\cite{Fradkin_1991}

\section{Single-particle scattering rate in a Fermi liquid due to Coulomb and electron-phonon interactions}
\label{app:B}

As a further illustration of our arguments leading to an effective action for calculating relaxation rates, let us consider
the well-known case of quasiparticle relaxation due to density fluctuations. To this end, we
consider the very simple case of spinless, noninteracting electrons with action $S_0$, and add a statically screened
Coulomb interaction 
\be
S_{\text{int}} = \int_k \delta n(k)\,v_{\text{sc}}({\bm k})\,\delta n(-k)\ .
\label{eq:B.1}
\ee
Here $v_{\text{sc}}({\bm k}) = 4\pi e^2/({\bm k}^2 + \kappa^2)$, with $\kappa$ the screening wave number,
and $n(k)$ is the Fourier transform of the electron number density $n(x) = {\bar\psi}(x)\psi(x)$. A finite average
density is already built into $S_0$ via the chemical potential, so $S_0$ serves the purpose of the reference
ensemble action $S_{\lambda}$ in Sec. \ref{sec:II} or Appendix \ref{app:A}. Now we follow the logic of 
Sec.\ \ref{subsec:II.A}. A number density fluctuation $\delta N$ will couple to the field $\delta n(x)$ via the
interaction $v_{\text{sc}}$ to produce an action
\bea
S[{\bar\psi},\psi] &=& S_0[{\bar\psi},\psi] 
\nonumber\\
&&\hskip -30pt + \int dx\,dy\ \delta N(x)\,\delta(\tau_x-\tau_y)\,v_{\text{sc}}({\bm x}-{\bm y})\,\delta n(y)\ ,\qquad
\label{eq:B.2}
\eea
and the density fluctuations are governed by a Gaussian action
\be
S_{\text{fluct}}[\delta N] = \frac{-1}{2}\int dx\,dy\ \delta N(x)\,\chi^{-1}(x-y)\,\delta N(y)\ ,
\label{eq:B.3}
\ee
with $\chi$ the physical density susceptibility. Integrating out the density fluctuations, we obtain an effective
action
\bse
\label{eqs:B.4}
\be
S_{\text{eff}}[{\bar\psi},\psi] = S_0[{\bar\psi},\psi] + \frac{1}{2} \int_k \delta n(k)\,V(k)\,\delta n(-k)
\label{eq:B.4a}
\ee
with an effective potential
\be
V(k) = (v_{\text{sc}}({\bm k}))^2\,\chi(k)\ .
\label{eq:B.4b}
\ee
\ese
Now we calculate the single-particle relaxation rate as in Sec.\ \ref{sec:III}. We obtain
\bse
\label{eqs:B.5}
\be
\frac{1}{2\tau} = \Gamma(\epsilon=0) = 2\NF\int_{-\infty}^{\infty} du\ {\bar V}''(u)\,\frac{1}{\sinh(u/T)}\ ,
\label{eq:B.5a}
\ee
where
\be
{\bar V}''(u) = \frac{1}{(\NF V)^2}\sum_{{\bm k},{\bm p}} \delta(\xi_{\bm k})\,\delta(\xi_{\bm p})\,V''({\bm k}-{\bm p},u)\ .
\label{eq:B.5b}
\ee
\ese
From Eq.\ (\ref{eq:B.4b}) we see that the spectrum of the potential $V$ is given by the spectrum of the density
susceptibility, which to lowest order in the screened Coulomb interaction is just the Lindhard function $\chi_0$.
For $\vert u\vert < (2\kF\vert{\bm k}\vert - {\bm k}^2)/2\me$, the spectrum of the latter is
\be
\chi_0''({\bm k},u) = \pi\NF\,u/\vF\vert{\bm k}\vert\ ,
\label{eq:B.6}
\ee
with $\vF$ the Fermi velocity. For the relaxation rate due to the electron-electron interaction we thus obtain the 
well-known Fermi-liquid result
\be
\frac{1}{2\tau_{\text{e-e}}} = \frac{\pi}{4}\,\frac{T^2}{\epsilonF}\ .
\label{eq:B.7}
\ee

The above derivation is similar in spirit to the arguments given in Ref.\ \onlinecite{Anderson_1984}. The point 
of this exercise is  to demonstrate that our heuristic method of coupling density fluctuations to
the appropriate fermion fields that we employed in Sec.\ \ref{sec:II} still works in this case where the 
density fluctuations are produced by the
very electrons they couple to. To put the result for the effective interaction, Eq.\ (\ref{eq:B.4a}), in context,
consider a bare Coulomb interaction, $v_{\text c}({\bm k}) = 4\pi e^2/{\bm k}^2$, and perform an RPA
resumption to produce a dynamically screened Coulomb interaction
\be
V_{\text{sc}}(k) = \frac{v_{\text c}({\bm k})}{1 + v_{\text c}({\bm k})\chi_0(k)}\ .
\label{eq:B.8}
\ee
To linear order in the frequency, the spectrum of the effective potential $V$ coincides with the spectrum
of $V_{\text{sc}}$, and $V$ therefore suffices to produce the leading low-temperature dependence of
the relaxation rate. Our effective action thus captures the leading effects of the soft modes in the system
(here, the soft particle-hole excitations that are reflected in the spectrum of the Lindhard function; in
Secs.\ \ref{sec:II}, \ref{sec:III}, the magnons). Note that it does {\em not} suffice to produce static screening,
which requires taking into account massive modes, which is why the above argument starts with a
statically screened interaction. Also note that the effective interaction $V$ is quadratic in the bare
interaction $v_{\text{sc}}$, in accordance with Fermi's golden rule. Analogously, the effective interaction
in Sec.\ \ref{sec:II}, Eq.\ (\ref{eq:2.12b}), is quadratic in the coupling constant $\Gammat$. We also mention
that the $T^2$ result, Eq.\ (\ref{eq:B.7}), holds for any short-ranged interaction, with the prefactor proportional
to the potential strength squared.

The above considerations assumed an electronic density fluctuation $\delta n$ interacting with a
density fluctuation $\delta N$ created by all other electrons, in analogy with magnetization fluctuations
in the case of an itinerant magnet. However, there is no reason why $\delta N$ cannot be a density
fluctuation extraneous to the electron system, in analogy to magnetization fluctuations due to electrons
in a band other than the conduction band. For instance, if $\delta N$ is an ionic density fluctuation, it will still
couple to $\delta n$ via a statically screened Coulomb interaction. Equations (\ref{eq:B.2}) - (\ref{eqs:B.5})
remain formally valid, except that the susceptibility $\chi$ now describes ionic density fluctuations, i.e.,
phonons. If we consider longitudinal phonons the susceptibility is the same as in a fluid and given by\cite{Forster_1975}
\be
\chi''({\bm k},u) = \pi\,\rho^2\,\kappa\,u^2\,\delta(u^2 - \omega^2_{\text{L}}({\bm k})) \ ,
\label{eq:B.9}
\ee
with $\omega_{\text{L}}({\bm k}) = c\vert{\bm k}\vert$ the longitudinal phonon frequency. Here $\rho$ is the
ionic number density, $c$ is the longitudinal speed of sound, and $\kappa = -(\partial V/\partial p)/V$, with
$V$ the system volume and $p$ the pressure, is the compressibility. We thus have
\be
{\bar V}''(u) = \frac{\pi}{16}\,\frac{\rho^2\kappa}{c^2 \kF^2 \NF^2}\, u^2\,\sgn u\ ,
\label{eq:B.10}
\ee
which leads to the familiar $T^3$ result for the single-particle scattering due to the
electron-phonon interaction in metals, 
\be
1/\tau_{\text{e-ph}} = \frac{7\pi}{3}\,\zeta(3)\,\frac{\rho^2\kappa}{\ne\,\me\, c^2}\,T^3\ .
\label{eq:B.10}
\ee


\begin{thebibliography}{55}
\bibitem[{D_f()}]{D_footnote}
\bibinfo{note}{If such a spin model is coupled to conduction electrons, the
  magnetization will still produce a splitting of the conduction band by virtue
  of the Zeeman effect, and the splitting will go to zero if and only if the
  magnetization goes to zero. $D$ can therefore still be considered a function
  of the splitting energy $\lambda$, even if the magnetization is caused by
  localized electrons in a different band.}
\end{thebibliography}

\end{document}